\documentclass[prl,aps,twocolumn,showpacs,preprintnumbers,amsmath,amssymb,twoside]{revtex4-1}

\usepackage{mathptmx}
\usepackage{graphicx}		% Include figure files
\usepackage{pstricks}
\usepackage{dcolumn}		% Align table columns on decimal point
\usepackage{bm}			% bold math
\usepackage{color}
\usepackage{units}		% Formating number and unit
\usepackage[dvipdfmx]{hyperref}
\usepackage{ulem} 		% For using \sout striking out command
\hypersetup{colorlinks=true,urlcolor=blue,linkcolor=blue,citecolor=blue}

\newcommand{\cco}{CaCu$_2$O$_3$}

\newcommand{\cle}{Cu $L_3$ edge}
\newcommand{\clr}{Cu $L_3$-edge RIXS}
\newcommand{\vb}[1]{\textcolor{black}{#1}}

\newcommand{\vbb}[1]{\textcolor{black}{#1}}

\begin{document}

\title{Femtosecond dynamics of magnetic excitations from resonant inelastic x-ray scattering in \cco}

\author{Valentina Bisogni$^{1,2}$}
\altaffiliation[Present address: ]{Research Department Synchrotron Radiation and Nanotechnology, Paul Scherrer Institut, CH-5232 Villigen PSI, Switzerland}
\email{valentina.bisogni@psi.ch}
\author{Stefanos Kourtis$^{1}$}

\author{Claude Monney$^{2,3}$}
\altaffiliation[Present address: ]{Fritz-Haber-Institut der Max-Planck-Gesellschaft, Faradayweg 4-6, 14195 Berlin, Germany}
\author{Kejin Zhou$^{2}$}
\altaffiliation[Present address: ]{Diamond Light Source, Harwell Science and Innovation Campus, Didcot, Oxfordshire OX11 0DE, UK}
\author{Roberto Kraus$^{1}$}
\author{Chinnathambi Sekar$^{1}$}
\altaffiliation[Present address: ]{Department of Bioelectronics and Biosensors, Alagappa University,Karaikudi-630 003, Tamilnadu, India}
\author{Vladimir Strocov$^{2}$}
\author{Bernd B\"{u}chner$^{1,4}$}
\author{Jeroen van den Brink$^{1,4}$}
\author{Lucio Braicovich$^{5}$}
\author{Thorsten Schmitt$^{2}$}
\author{Maria Daghofer$^{1}$}
\author{Jochen Geck$^{1}$}
\affiliation{$^1$Leibniz Institute for Solid State and Materials Research IFW Dresden, 01069 Dresden, Germany\\
$^2$Research Department Synchrotron Radiation and Nanotechnology, Paul Scherrer Institut, CH-5232 Villigen PSI, Switzerland\\
$^3$Fritz-Haber-Institut der Max-Planck-Gesellschaft, Faradayweg 4-6, 14195 Berlin, Germany\\
$^4$Department of Physics, Technical University Dresden, D-1062 Dresden, Germany\\
$^5$ CNR/SPIN, CNISM and Dipartimento di Fisica, Politecnico di Milano,
Piazza Leonardo da Vinci 32, 20133 Milano, Italy}

\date{Received: \today}

\begin{abstract}
Taking spinon excitations in the quantum antiferromagnet \cco~ as an example, we demonstrate that femtosecond dynamics of magnetic excitations can be probed by direct resonant inelastic x-ray scattering (RIXS). To this end, we isolate the contributions of single and double spin-flip excitations in experimental RIXS spectra, identify the physical mechanisms that cause them and determine their respective timescales. By comparing theory and experiment, we find that double spin flips need a finite amount of time to be generated, rendering them sensitive to the core-hole lifetime, whereas single spin flips are to a very good approximation independent of it. This shows that RIXS can grant access to time-domain dynamics of excitations and illustrates how RIXS experiments can distinguish between excitations in correlated electron systems based on their different time dependence.
\end{abstract}


% 75.30.Ds	Spin waves
% 78.70.Ck	X-ray scattering
% 78.47.-p	Spectroscopy of solid state dynamics
% 78.47.J-	Ultrafast spectroscopy (<1 psec)

\maketitle

Traditional spectroscopic methods used in condensed matter experiments relate dynamical properties of a material to frequency-dependent correlation functions. % of the constituent particles. 
These techniques have been very successful in probing elementary excitations of solids, revealing their sometimes unconventional nature. For instance, using frequency-resolved spectroscopies to study low-dimensional cuprates, electronic quasiparticles that are fractions of an electron, in the sense that they carry either its spin, charge or orbital, have been detected \cite{Kim2006,Lake2009,Schlappa2012}. 
%
%The time-dependent behaviour of such excitations is a complementary aspect, which may help \vb{resolving} characteristics that cannot be disentangled in the frequency domain or even reveal unforeseen properties of matter~\cite{Tokura2006,Yonemitsu2006}.  
%

Technological advances in intense pulsed radiation sources, including those at novel x-ray free electron laser (XFEL) facilities~\cite{Emma2010}, nowadays enable to observe dynamics via time-resolved measurements. 
A usual scheme is to first excite a system using a pump pulse and then to probe the excited state at a later time by means of a probe pulse. Today, these pump-probe experiments allow studying the temporal evolution of excited states in the pico-, femto- and attosecond timescales~\cite{Bloembergen1999,Cavalieri2007a,Krausz2009}.
The pump-probe approach provides opportunities for further exploration of correlated electron materials, because the femtosecond to picosecond range corresponds to dynamics usually determined by electron-electron or electron-phonon interactions -- processes of key importance for the physical properties of these materials. % like e.g. the high-temperature superconductors~\cite{...}.
% Such fast dynamics can be matched by conventional pulsed or novel x-ray free electron laser (XFEL) facilities~\cite{Emma2010}, the latter combining the time-resolved pulsed mode with %the \vb{key} properties \vb{such as} % of x-rays, i.e. energy tuning and high brilliance \vb{of the x-rays}. 

An alterative route towards the time domain is core-level spectroscopy, either involving outgoing electrons \cite{vbBruhwiler2002} or photons \cite{Skytt1996,Foehlisch2006,Braicovich2008,Pietzsch2011}. Initially, a core electron is promoted into an unoccupied orbital and an intermediate perturbed state with a finite lifetime, the core-hole lifetime $\tau$, is created. 
At time $\tau$ after the core-hole creation, an electron decays into the core hole, leading to emission of either a photoelectron or an x-ray photon. The measured signal carries information on the generation and propagation of excitations created during the core-hole lifetime.
In the case of $L-$edges of $3d$ transition metals, the $2p$ core-hole lifetime turns out to be in the femtosecond range, corresponding to the dynamics of correlated electrons.

Resonant inelastic x-ray scattering (RIXS) is a {\it photon in -- photon out} technique, which involves the creation of a core-hole in the intermediate state and therefore implements the aforementioned core-hole clock. In addition, high-resolution RIXS is a unique technique for probing magnetic, orbital and charge excitations in strongly correlated materials~\cite{vbHill2008,vbBraicovich2009,vbSchlappa2009,vbBraicovich2010,vbGuarise2010,vbament2011}. The combination of these two properties therefore provides an approach to access the time evolution of electronic excitations, a feature which so far has only barely been exploited. This marks a substantial advantage when facing the problem of disentangling energy-overlapping excitations of distinct nature, for example charge and magnetic excitations in superconducting cuprates, by using their different time evolution in the intermediate state. 
%therefore provides an approach to access the time evolution of electronic excitations; a feature which so far, however, has only barely been exploited.  

In this Letter, we demonstrate how the temporal evolution of electronic excitations during the core-hole life time leaves its fingerprint on the RIXS cross section. As an example, we consider the elementary magnetic excitations of the spin chains in  \cco. Via Cu $L$-edge RIXS we measure the two types of magnetic excitations in this material, which correspond to a single spin-flip with a total spin change of $\Delta S=1$ ($S_1$) and a double spin-flip with $\Delta S=0$ ($S_0$)~\cite{vbBraicovich2010, vbSchlappa2009,vbament2009,vbKourtis2012,vbBisogni2012}. 
%$S_0$ and $S_1$ excitations are obtained \vbb{by flipping one spin or two antiparallel spins, respectively.}  This has been demonstrated recently for the two-dimensional cuprates, where these two types of excitations and their contribution as a function of energy loss and line shape has been studied in detail~\cite{vbBraicovich2010,vbGuarise2010,vbBisogni2012}. 
Following a previously established approach~\cite{vbBisogni2012}, we then extract the $S_0$ and $S_1$ contributions from the experimental spectra and compare the result to the theoretical RIXS response \vbb{obtained} by an exact-diagonalization
simulation of the time dependence of the RIXS process in a 1D $S=1/2$ chain.
We show that the $S_0$ and $S_1$ excitations evolve on very different time scales and therefore contribute differently to the measured RIXS signal. Specifically, we establish that the $S_1$ part, which depends on strong spin-orbit coupling, is created mainly upon decay and is affected only slightly by the core-hole lifetime, whereas the $S_0$ part takes a significantly long time to be created and depends crucially on intermediate-state dynamics: the intrinsic time constant of the $S_0$ excitation is $\tau_{S_0}=\hbar/J$, where $J$ is the spin-spin interaction strength. The ratio of core-hole lifetime $\tau$ to $\tau_{S_0}$ determines the $S_0$ contribution to the spectra.
 
\begin{figure}[t!]
\center{
\resizebox{0.95\columnwidth}{!}{%
 \includegraphics[clip,angle=0]{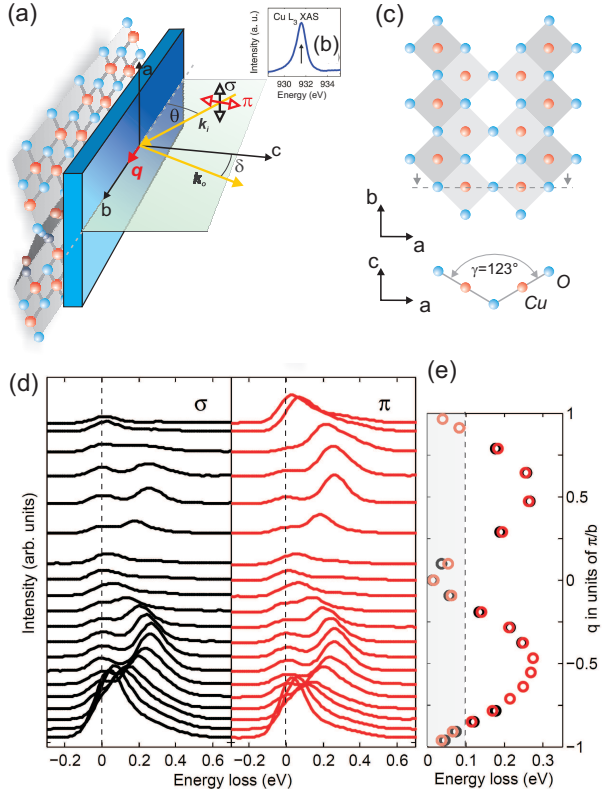}}}
\caption{(Color online) (a) Experimental layout for the RIXS measurements. The scattering angle was set to 130$^{\circ}$. (b) X-ray absorption at \cle~of \cco. (c) \cco~ structure in the $ab$ and $ac$ planes. (d) RIXS spectra of  \cco~for $\sigma$ (black) and $\pi$ (red) incoming polarized light, after being normalized to the $dd$-area, between 1.3 eV and 2.8 eV (not shown here). The vertical offset between two consecutive spectra is proportional to the difference of the corresponding momenta [given by the $y$ scale of panel (e)]. (e) Dispersion of the magnetic peak for the two polarizations (same color code). All the data have been measured at 40~K, above the N\'{e}el transition of \cco~ ($T_N=25$~K).}
\label{fig1}
\end{figure}

The RIXS experiment has been performed at the ADRESS beamline of the Swiss Light Source at the Paul Scherrer Institut
\cite{vbStrocov2010}, using the SAXES spectrometer \cite{vbsaxes}. The experimental layout \vbb{is illustrated} in Fig.~\ref{fig1}(a). The energy of the incoming x-ray photons was tuned to the maximum of the Cu $L_3$ x-ray absorption peak at 931.5~eV [cf. Fig.~\ref{fig1}(b)], \vb{for both linear {\it in-plane} ($\pi$) and {\it out-of-plane} ($\sigma$) polarizations $\epsilon$.} The combined energy resolution at this edge was 130~meV.

\cco~single crystals were grown as described in Ref.\,\cite{vbsekar2005}. This material realizes a two-leg spin-1/2 ladder, as illustrated in  Fig.\ref{fig1}(c). An important feature of this particular ladder structure is the 123$^\circ$ bond angle along the rung direction $a$, which results in anisotropic exchange constants $J_{\textrm{leg}}\simeq 10 \times J_{\textrm{rung}}\simeq 140-160$\,meV~\cite{vbbordas2005, Lake2009}. The 2-leg ladders can therefore be approximated as two weakly coupled spin-1/2 chains, since the effects of the coupling along the rung direction only become important at excitation energies below 10~meV~\cite{Lake2009}.

% in this system. 

%Wide energy range RIXS spectra of \cco~ are shown in Fig.~\ref{fig1}d, for incoming $\epsilon=\sigma$ (black line) and $\epsilon=\pi$ (red line) polarized light and for $q=-0.47$ in units of $\pi/b$, where $b=4.1$~\AA. As in the case of \clr~ spectra of two-dimensional cuprates~\cite{vbMoretti2011}, we identify $dd$-excitations at around 2~eV of energy loss $\hbar\omega\equiv\hbar\omega_{i}-\hbar\omega_{o}$, where $\hbar\omega_i$ and $\hbar\omega_o$ the incoming and outgoing beam energies respectively, weak charge transfer excitations extending up to 8~eV and magnetic excitations below 0.5~eV, clearly distinguishable from the quasi-elastic peak. 

%In this Letter
\vbb{In the following}, we focus on the magnetic RIXS excitations at energies between 100 and 300\,meV where the effects of $J_{\textrm{rung}}$ play no role.
The corresponding experimental RIXS spectra for different momentum transfers $q$ are shown in Fig.~\ref{fig1}(d) as a vertical stack with $q$ ranging between -1 to 1 (in units of $\pi/b$, where $b=$4.1\AA) and for $\sigma$- (black) and $\pi$- (red) incoming polarized light. \vb{Already from the raw data the dispersion of the magnetic excitations is very clear.} As can be seen in Fig.~\ref{fig1}(e), the peak position tracks the lower bound of the two-spinon excitations and reaches a maximum at $q\simeq$0.5. In addition to this, our RIXS data reveals the continuum of 2-spinon excitations at higher energies above the lower bound. This is exactly what is expected for a spin-1/2 chain~\cite{vbZaliznyak2004} and agrees very well with previous inelastic neutron scattering data~\cite{Kiryukhin2001,Lake2009}. Thus, we ascertain that, \vb{within the energy regime accessed by the present RIXS experiment}, the magnetic excitations of \cco~ correspond to 2-spinon excitations of AFM spin-1/2 chains with a superexchange $J\approx$160 meV. %~\cite{Kiryukhin2001,Lake2009}. 

% The intensity of the magnetic excitation in RIXS is polarization
% $\epsilon$ dependent. Also, it is not depending on the momentum
% transferred $|\q|$; instead it depends on the experimental geometry
% $\pm$\q. The behavior of the RIXS intensity, therefore, is  mainly
% determined by the local symmetry of the considered excitation
% \cite{vbament2009,Braicovich2010a,Haverkort2010}. Qualitatively,
% similar behavior of the RIXS magnetic intensity has been found in
% the two-dimensional cuprates and attributed to the $S_1$ component
% (namely the single magnon) which is dominating the spectral weight
% in the low energy region \cite{Braicovich2010a}. Being the local
% \cuo~configuration and the scattering condition similar in the two
% cases, the mentioned correspondence supports a dominant component of
% $S_1$ character in the \cco~ two-spinon excitation.  
% 
% Furthermore, it is also reasonable to expect a $S_0$ two-spinon
% contribution from the present data. Following the analogy with 2D
% cuprates, it has been in fact observed that RIXS at \cle~ can
% additionally create a $S_0$ magnetic excitations (namely, bimagnons)
% with smaller weight than the $S_1$ contribution \cite{Braicovich2010,
% Bisogni2012}.  

\begin{figure}[h!]
\center{
\includegraphics[width=.95\columnwidth]{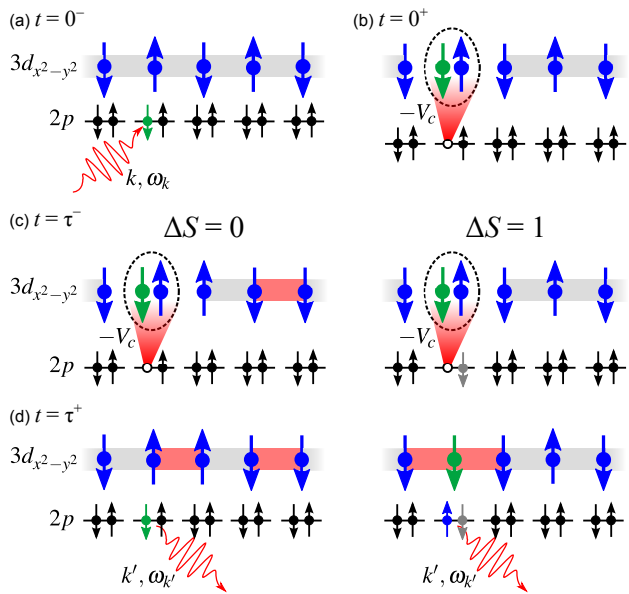}}
\caption{(Color online) Direct magnetic RIXS processes. (a) Schematic
  representation of the initial state $2p^63d^{9}$, i.e. the ground state of a 1D AFM $S=1/2$ chain, at time $t=0^-$ just before the creation of the
  core hole. (b) Intermediate state with one doublon and one
  core hole present ($2p^53d^{10}$ configuration) right after the
  creation of the core hole at $t=0^+$. As the $3d$ level at the core-hole
  site is doubly occupied and has a net spin of 0, magnetic coupling
  to neighboring sites vanishes. (c) Intermediate state at a later
  time $t=\tau^-$, just before the de-excitation of the core hole. On
  the left side, two spins next to the core-hole site have flipped due to scattering on the doubly occupied site, i.e., the ``cut''
  in the spin chain. On the right
  side, spin-orbit coupling of the $2p$ state has flipped the
  core-hole spin. (d) Two-spinon excitations (illustrated by darker
  bonds) in the final state at
  $t=\tau^+$. In (c) and (d), the left (right) part illustrates the
  $S_0$ ($S_1$) process.}  
\label{fig2}
\end{figure}

Figure~\ref{fig2} provides a schematic illustration of the dominant mechanisms, by which \clr~ creates 2-spinon excitations in an AFM spin-1/2 chain. Initially at $t=0$, a $2p$ core electron is excited into the $3d$ valence shell [see Fig.~\ref{fig2}(b)].
%; it drops back to the core level in the final step (d). 
%
During the lifetime $\tau$ of the created core hole, two different processes can occur that leave magnetic excitations behind: (i) due to spin-orbit coupling in the $2p$-state the core hole spin can flip [Fig.~\ref{fig2}(c), right panel]. After the core-hole decay, this results in a local spin-flip in the valence shell of the excited site  [Fig.~\ref{fig2}(d), right panel], yielding the $S_1$ excitation. (ii) Two spins next to the core-hole site scatter at the doubly occupied site [Fig.~\ref{fig2}(c), left panel]. After the decay of the core-hole state again 2 spinons have been created [Fig.~\ref{fig2}(d), left panel], but this time the spinons have opposite spins and hence the total spin did not change. This is the $S_0$ process. 
%%%%%%%%%%%%%%%%%%%%%%%%%%%%%%
% \textcolor{blue}{MD: This info should also be here, but I am not sure where to put it into the paragraph:}
% should be mentioned later on
%jg: Aren't we making the point that the different intensities are due to the different timescales?! This is now confusing

%Due to its non-perturbative nature, the single spin-flip $S_1$ process is dominant, whereas the double spin-flip $S_0$ gives a weaker contribution to the cross section and other higher order processes can be neglected~\cite{vbKourtis2012,vbIgarashi2012}. 
%expected to be much more prominent in the spectra than double \vbb{spin} flip
%processes~\cite{vbKourtis2012,vbIgarashi2012}, and multiple \vbb{spin} flip processes have vanishing weight in both channels. 
% {[Comment: it sounds a bit of a repetition]} 
% Double-flips in addition to the single-spin flip also occur in the $S_1$ channel, but as the single flip dominates, their impact is subtle and small~\cite{vbKourtis2012,vbIgarashi2012}. 
%Only in the $S_0$ channel, where the single \vbb{spin} flip is absent, do double  \vbb{spin} flips become crucial. 
%%%%%%%%%%%%%%%%%

\vb{For a specific polarization $\epsilon$} the magnetic RIXS intensity can be expressed formally as $I^\epsilon(q,\omega,\tau) = I^\epsilon_{S_0}(q,\omega,\tau) + I^\epsilon_{S_1}(q,\omega,\tau)$. To quantify these observations, we use the factorization introduced in Refs.~\onlinecite{vbament2009,Haverkort2010}, which can be written as $I^\epsilon(q,\omega,\tau) = F^\epsilon_{S_0}(q)\cdot G_{S_0}(|q|,\omega,\tau)$ + $F^\epsilon_{S_1}(q)\cdot G_{S_1}(|q|,\omega,\tau)$, where $F$ are local form factors depending on the geometrical parameters $\epsilon$, $q$ and corresponding to the transition between spin-orbit split $2p_{3/2}$ states to $3d_{x^2-y^2}$ state, while $G$ are dynamical structure factors depending on energy transfer $\omega$, magnitude of momentum transfer $|q|$ and -- most importantly for the
present analysis -- core-hole lifetime $\tau$. 
Following Refs.~\onlinecite{vbament2009,vbBraicovich2010a}, $F^\epsilon _{S_1}$ is replaced by the local spin-flip probability
$P_{\textrm{sf}}^\epsilon$. Additionally, we assign to $F^\epsilon_{S_0}$ the elastic probability $P_{\textrm{el}}^\epsilon$,
because the state of the excited site has not changed during the $S_0$ process.
% according to the discussion of  $S_0$ process given above. %the preceding paragraph and Fig.~\ref{fig2}(c). 
The overall magnetic RIXS intensity is thus given by 
\begin{equation}
I^\epsilon(q,\omega,\tau)= P_{\textrm{el}}^\epsilon(q) \cdot G_{S_0}(|q|,\omega,\tau) + P_{\textrm{sf}}^\epsilon(q) \cdot G_{S_1}(|q|,\omega,\tau) .
\label{eq1}
\end{equation}
The probabilities $P_{\textrm{el}}^\epsilon(q)$ and $P_{\textrm{sf}}^\epsilon(q)$ can be evaluated within the single ion model~\cite{vbMoretti2011} as described in the Supplementary Material (SM). 
Considering a set of two equations (\ref{eq1}) for $\epsilon=\sigma$ and $\epsilon=\pi$ at a given  $q$, it is possible to extract the two unknown quantities $G_{S_0}(|q|,\omega,\tau)$ and $G_{S_1}(|q|,\omega,\tau)$ as a function of  $P_{\textrm{el}}^\epsilon(q)$,
$P_{\textrm{sf}}^\epsilon(q)$  and the measured magnetic RIXS intensities $I^\sigma(q,\omega,\tau)$ and $I^\pi(q,\omega,\tau)$ (see Ref.~\onlinecite{vbBisogni2012} and SM). This procedure thus allows, firstly, to determine $G_{S_0}(|q|,\omega,\tau)$ and $G_{S_1}(|q|,\omega,\tau)$, which both do not depend on $\epsilon$, and secondly, to extract $I^\epsilon_{S_0}(q,\omega,\tau)$  and $I^\epsilon_{S_1}(q,\omega,\tau)$.
%Repeating this procedure for each measured $q$, the momentum dependence of these quantities is obtained. 

%Then, using the procedure described in Ref.~\onlinecite{vbBisogni2012} and the SM, the $S_0$ and $S_1$ contributions can be extracted from the experimental data for each measured $q$, which allowsfor the determination of $G_{S_0}$ and  $G_{S_0}$ as a function of $q$. %the line shapes of  \vb{the dynamical structure factors} %, as done in  Refs.}~\onlinecite{vbBraicovich2010,vbBisogni2012}, including additionally  and their momentum dependence. 

\begin{figure}[t!]
\center{
\includegraphics[width=0.98\columnwidth]{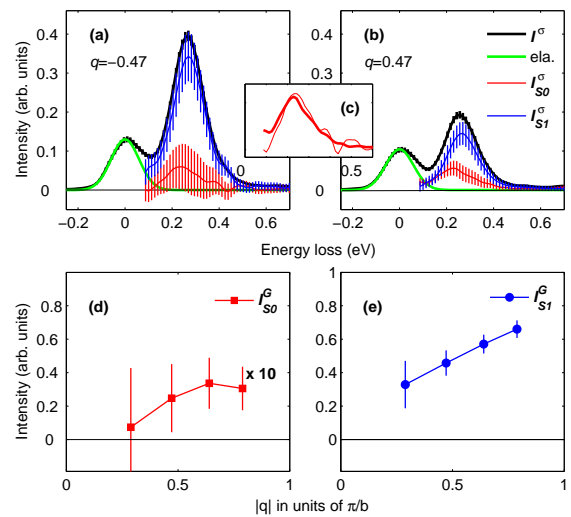}}
\caption{(Color online) (a), (b): Spectral decomposition of total RIXS intensity $I^\sigma$ (black dotted line) into $S_0$ (red line) and $S_1$ (blue line) components for $q=-0.47$ and $q=0.47$. The elastic signal is shown as green line. (c) Comparison between the extracted $I_{S_0}$ spectral shapes at $q=-0.47$ (thin line) and $q=0.47$ (thick line). (d) $I^G_{S_0}$ and (e) $I^G_{S_1}$ as a function of $q$, estimated by integrating $G_{S_0}$ and $G_{S_1}$ between 0.1 and 0.6~eV. Only points corresponding to positive $q$ are considered (see SM for more details).}
\label{fig3}
\end{figure}

%While details of this analysis are presented and explained in the SM, here we directly introduce and discuss the results. 
Representative results of this analysis are given in Fig.~\ref{fig3}. In panels (a) and (b), RIXS spectra taken at $q=-0.47$ and $q=0.47$ with $\epsilon=\sigma$ are shown as well as their decomposition into $I^\sigma_{S_1}$ (blue) and  $I^\sigma_{S_0}$ (red). The extracted $S_0$ and $S_1$ contributions at a given $q$ show a large energy overlap, in qualitative agreement with theoretical expectations~\cite{vbKourtis2012, vbIgarashi2012, klauser2011}.
By comparing the  independent results for the two $q$-values, it can be seen that the energy position of each magnetic component coincides for $\pm q$, as demonstrated in Fig.~\ref{fig3}(c) for the $S_0$ case. This strongly supports the validity of our analysis, because it correctly yields  $G_{S_0,S_1}=G_{S_0,S_1}(|q|)$, without incorporating this as a condition into the decomposition.
%
%where the maxima of both $I^\sigma_{S_0}(+q)$ and $I^\sigma_{S_0}(-q)$ are found at $\simeq 250$~meV. 
%
 
%As far as it concerns the intensity, instead, the difference (similarity) in the $I^\sigma_{S_1}$ ($I^\sigma_{S_0}$) spectral weight between $q=-0.47$ and $q=0.47$ in Figs.~\ref{fig3}(a-b) is consistent with the respective difference (similarity) in the $P_{\textrm{sf}}^\sigma$ ($P_{\textrm{el}}^\sigma$) factors (see SM) at these $q$ points.

As described above, the analysis also yields the dynamical structure factors $G_{S_0}$ and $G_{S_1}$. Their integrated spectral weights as a function of $|q|$, $I^G_{S_0}$ and $I^G_{S_0}$, are shown in Figs.\,\ref{fig3} (d) and (e), respectively.
%
%Knowing the $I^\epsilon_{S_0}$ and the $I^\epsilon_{S_1}$ components at each $q$, equivalently corresponds to know the respective dynamical RIXS responses,
%$G_{S_0}=I^\epsilon_{S_0}/P^\epsilon _{el}$ and
%$G_{S_1}=I^\epsilon_{S_1}/P^\epsilon_{sf}$. The information on $G$ is
%here presented as the total magnetic spectral weights $I^G_{S_0}$ and
%$I^G_{S_1}$ as a function of $q$ [see Figs.~\ref{fig3}(d) and (e)]. 
%
Interestingly, the integrated spectral weights clearly display different behaviors: while that of the $S_0$ component has a broad maximum around $q=0.6$, the $S_1$ part increases steadily with $q$. Furthermore, depending on $q$, the $S_0$ contribution is roughly 10 to 20 times weaker in strength than $S_1$.
%, in qualitative agreement with what discussed before during the explanation of the two processes. 

%\vb{While} the details of the analysis are presented and explained in the SM, here we focus on the central outcome of this procedure.Figs.~\ref{fig3}(a-b) show RIXS spectra taken at $q=-0.47$ and $q=0.47$ for $\sigma$-polarized %rays\vb{photons}, decomposed into $I^\sigma_{S_1}$ (blue), $I^\sigma_{S_0}$ (red) and the elastic peak (green). The maxima of the two inelastic components almost coincide, within experimental accuracy, at $\hbar\omega\simeq 270$~meV. %

%MOVE THIS SENTENCE TO SM	
%The widths are comparable as well: being the full widths at half maximum 180~meV, this indicates that the excitations are intrinsically broader than the experimental resolution (130~meV). 

%The difference (similarity) in the $I^\sigma_{S_1}$ ($I^\sigma_{S_0}$) intensity between $q=-0.47$ and $q=0.47$ in Figs.~\ref{fig3}(a-b) is consistent with the respective difference (similarity) in the $P_{\textrm{sf}}^\sigma$ ($P_{\textrm{el}}^\sigma$) factors (see SM).

% We also notice that the line shapes of $I^\sigma_{S_0}$ for $q=-0.47$ and $q=0.47$ coincide very well with each other, as shown in Fig.~\ref{fig3}(c). This is indeed an important indicator for the correctness of the analysis, showing that $G_{S_0}$ is  only a function of $|q|$.

We now turn to the theoretical modeling of the time evolution of magnetic excitations in a spin-1/2 chain and its relation to the RIXS cross section. 
%\vb{Theoretically}, the \tbr{quench defined by} 
The effect of the core-hole lifetime $\tau$ on the RIXS process was simulated using exact diagonalization~\cite{vbKourtis2012} for a ring
of up to 16 sites modeling the $3d$ valence states and a
time-dependent term describing the core hole:
\begin{equation}
 {\mathcal H}(t) = {\mathcal H}_{3d} + \Theta(t)\Theta(\tau-t) {\mathcal H}_{\textrm{corehole}}\;.
\end{equation}
${\mathcal H}_{\textrm{corehole}}$ consists of a local attractive
potential at a single site, which is switched on (at $t=0$) and off (at $t=\tau$)
together with the creation and annihilation of the extra $d$ electron, respectively, see the illustration in Fig.~\ref{fig2}. The effective spin-flip in the $S_1$ process, see Fig.~\ref{fig2}(c), is incorporated by creating and annihilating electrons with opposite spin in the $3d_{x^2-y^2}$ orbital at the core-hole site. Note that the core-hole lifetime $\tau=\hbar/\Gamma$, $\Gamma$ being the core-hole lifetime broadening,  is explicitly taken into account in both the $S_0$ and $S_1$ cases. 

We compared several microscopic models of ${\cal H}_{3d}$, modeling the
$3d_{x^2-y^2}$ chain with Hubbard, $t$-$J$ and Heisenberg models. In the
former two, the doubly occupied site (doublon) is in principle allowed to move away from the core hole during the RIXS process, but for the experimentally determined core-hole potential in cuprate materials of 9 eV~\cite{Vanveenendaal1993}, the doublon hopping was found to play no role. 
%this does not affect the results. 
The relevant aspect for creating the $S_0$ excitations
thus turns out to be the vanishing spin of the doubly occupied
site, which cuts the $S=1/2$ chain. In the following, we will therefore
present results obtained by using a Heisenberg Hamiltonian, whose only
energy scale $J$ defines the spinon-propagation timescale $\tau_{S_0}=\hbar/J \approx 4 fs$ ($J=0.16$ eV for \cco).

\begin{figure}[t!]
\center{
\includegraphics[width=0.98\columnwidth]{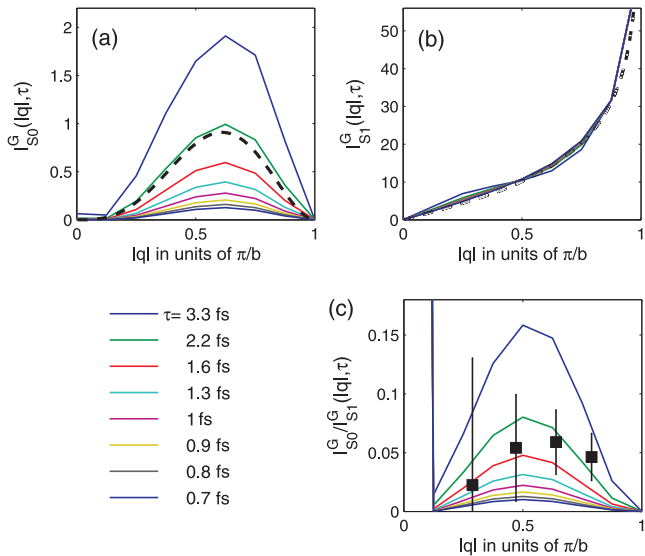}}
\caption{(Color online) Theoretical magnetic RIXS response and comparison to experiment. Numerically obtained (a) $I^G_{S_0}(|q|)$ and (b) $I^G_{S_1}(|q|)$ as a function of $\tau$, calculated with exact diagonalization (solid colored lines), and comparison to theoretical $I^G_{S_0}(|q|)$ (rescaled) from numerical Bethe ansatz calculations~\cite{klauser2011} (dashed black lines). (c) $(I^G_{S_0}/I^G_{S_1})(|q|)$ as a function of $\tau$ compared to experimental data points (black squares). Calculations have been done for $J=0.16$~eV.}
\label{fig4}
\end{figure}

\vb{From Figs.~\ref{fig4}(a-b)}, it can be seen that the
theoretical results \vb{for $I^G_{S_0}$ and $I^G_{S_1}$} closely follow the dispersive trends of the experimental findings shown in Figs.~\ref{fig3}(d-e). Comparison to Bethe Ansatz results from Ref.~\onlinecite{klauser2011} (black dashed lines in Figs.~\ref{fig4}(a-b)) shows that finite-size effects are small. \vb{By increasing the lifetime $\tau$}, the calculated $I^G_{S_0}$ %is expected to 
strongly increases [Fig.~\ref{fig4}(a)], while the $I^G_{S_1}$ remains almost unchanged~\cite{vbKourtis2012,vbIgarashi2012} [Fig.~\ref{fig4}(b)]. This behavior is expected, since the spin flip in the $S_1$ process is directly introduced upon the decay and therefore dominates the response, while perturbative effects due to the intermediate-state doublon contribute only minutely to $I^G_{S_1}$.  
 %since the spin flip in the $S_1$ process is non-perturbative and dominates the response, while perturbative effects due to the intermediate-state doublon contribute only minutely to the spectra. 
On the contrary, $S_0$ excitations are created only due to the presence of the intermediate-state doublon. During the timespan $\tau$, excitations are generated at and propagate away from the core-hole site. $I^G_{S_0}$ therefore increases strongly with increasing $\tau$ and becomes significant only for $\tau\sim\hbar/J$. $S_1$ excitations, on the other hand, can occur even for $\tau\ll\hbar/J$.

For a quantitative comparison \vb{between theory and experiment}, we \vb{introduce the ratio $(I^G_{S_0}/I^G_{S_1})(|q|,\tau)$}.
%, a key quantity to reveal the real-time dynamics of the $S_0$ excitations. This
%
This is shown in Fig.~\ref{fig4}(c), where the experimental data points (black squares)
are compared to the theoretical results for different $\tau$ (solid
lines). This comparison yields
% The value extracted from the \vb{best fitting curves} -- 
%
$\tau \approx $ 1.6 - 2.2~fs, in good agreement with experimental estimates of the core-hole lifetime in \clr~($\Gamma \sim$~0.3 - 0.4~eV) \cite{Nyholm1981,*Fuggle1992}. This result indicates that the generation of $S_0$ excitations in \cco~ is evidently a slow process compared to the core-hole
lifetime and that the intensity of $S_0$ excitations scales as the dimensionless timescale ratio $\tilde\tau=\tau/\tau_{S_0}=J/\Gamma$. The non-zero $S_0$ weight for $\tau<\tau_{S_0}$ is a quantum-mechanical effect.
 % as a unitless time measure of the effective duration of the \tbr{quench-like} $S_0$ process. 
 %
 
 \vbb{By tuning the $\tilde\tau$ knob, it is therefore possible to investigate the time dynamics of the $S_0$ excitation further. This can be experimentally achieved in several ways, by controlling independently $J$  - i.e. by considering different materials, isovalent dopings, or strain \cite{vbminola2012} - and $\Gamma$, i.e. by moving to another absorption edge.}
 % There are several ways to control \vb{the effective quench duration} $\tilde\tau$ and therefore investigate the real-time dynamics of $S_0$ excitations further. The most straightforward one is to consider another quasi-1D compound with larger $J$, like Sr$_2$CuO$_3$ for which $J\simeq 250$~meV~\cite{vbWalters2009}, while keeping the same excitation edge, i.e.\ the same $\tau$. As a result, the $(I^G_{S_0}/I^G_{S_1})(|q|,\tau)$ ratio should be approximately doubled, as shown in Fig.~\ref{fig4}(c) (dashed magenta line), meaning that double is the \vb{expected} intensity of the $S_0$ excitations in Sr$_2$CuO$_3$ compared to \cco, while the $S_1$ weight should remain almost unchanged. Other experimental knobs, offering finer control of $\tilde\tau$, could be isovalent doping or application of pressure to the sample \cite{vbminola2013}. Finally, another way to tune the quench duration would be to move to another absorption edge, thus changing the intrinsic $\tau$.
%\vbb{[Comment: very important concept to have also in the introduction.]}  These control parameters in the time domain may offer an enormous advantage when facing the problem of disentangling energy-overlapping excitations of different nature, such as charge and magnetic contributions in the superconducting cuprates. Even more importantly, the different timescales needed for generating different elementary excitations open the door to disentangle and identify them using time resolved RIXS experiments.
Future perspectives of RIXS experiments at XFEL facilities in stimulated-emission mode promise continuous control of the timescale ratio $\tilde\tau$, which may be used to disentangle excitations with different intrinsic time scales that overlap in the energy domain. For example, $S_0$ excitations in \cco~ could be suppressed by decreasing $\tilde\tau$.

% Mostly, the $S_0$ contribution is sensitive to the dynamics of the
% perturbation induced at the RIXS intermediate state. This can be
% explained because $S_0$ is a slow process requiring the interactions
% between spins, with a time scale $\tau_{sf}=\hbar/J\approx$ 4
% $fs$. Since this number is comparable with the core-hole lifetime
% $\hbar/\Gamma\approx$ 2 $fs$, therefore a finite $S_0$ contribution is
% present in the RIXS spectra. Consequently, for larger values of the
% $J/\Gamma$ ratios, the $S_0$ contribution is expected to be stronger. A
% practical example could be realized by considering systems with
% larger $J$, i.\,e. Sr$_{\rm 2}$CuO$_{\rm 3}$ ($J$=250 meV
% \cite{vbWalters2009}) while keeping the same excitation edge,
% i.e. the same $\Gamma$. As a result, the expected
% $I_{GS_0}^{th}(|\q|)$/$I_{GS_1}^{th}(|\q|)$ ratio doubles with respect
% to the present case, as shown in Fig.\,\ref{fig4}d, meaning that
% actually the $S_0$ dynamical structure factor is expected to double in
% Sr$_{\rm 2}$CuO$_{\rm 3}$ with respect to \cco, while the $S_1$
% dynamical  
% structure factor remains almost unchanged. 

In conclusion, we have experimentally determined the single ($S_1$) and double ($S_0$) spin-flip contributions to the measured magnetic RIXS spectra of \cco. We have then isolated the corresponding $S_1$ and $S_0$ \clr~ dynamical structure factors and have found that
their ratio is directly influenced by the spin dynamics within the
core-hole lifetime window, by comparing them to simulations of spin-excitation dynamics in a 1D $S=1/2$ AFM system. By combining experimental findings and
numerical results, we have shown that the $S_1$ process
is essentially insensitive to changes in all relevant timescales,
%(although we expect higher-edge RIXS to alter this simple picture)
\vbb{while the $S_0$ belongs to the femtosecond timescale and its response can be modified by changing the ratio $J/\Gamma$, experimentally feasible by changing, e.g., %one resembles a quench, whose duration can be controlled by 
stoichiometry, absorption edge or strain, or by using forthcoming XFEL sources.} %This conclusion opens the way to new possibilities of investigating magnetic excitations while tuning the $J/\Gamma$ knob. It should be noted that our analysis is not limited to magnetic excitations; it can be straightforwardly generalized to treat other well-defined excitations as well. 

%We anticipate that our analysis and results will be of particular relevance to forthcoming experiments in XFEL facilities, where control over the core-hole lifetime may be furnished by stimulated emission. 

{\it Acknowledgements --} We acknowledge fruitful discussions with K. Wohlfeld and S. Johnston. This work was performed at the ADRESS beamline using the SAXES instrument jointly built by Paul Scherrer Institut, Switzerland and Politecnico di Milano, Italy. \vbb{This project was supported by the Swiss National Science Foundation and its National Centre of Competence in Research MaNEP. The research leading to these results has received funding from the European Community's Seventh Framework Programme (FP7/2007-2013) under grant agreement Nr. 290605 (PSIFELLOW/COFUND).} We gratefully acknowledge the financial support through the Emmy-Noether Program (V.B., R.K. and J.G. under Grant GE1647/2-1, M.D. and S.K. under Grant DA 1235/1-1). V.B. also acknowledges the financial support from Deutscher Akademischer Austauschdienst (DAAD). 

\end{document}